\documentclass[a4,10pt]{article}
\pdfoutput=1
\usepackage[a4paper,hmargin={1.8cm,1.8cm},vmargin={2cm,2cm}]{geometry}
\usepackage{multicol} 

\usepackage[none]{hyphenat} 

\usepackage{gensymb}

\usepackage{cleveref}

\usepackage{graphicx}
\usepackage{float}
\usepackage[labelfont=bf]{caption}

\usepackage[square, numbers, comma, sort&compress]{natbib} 

\usepackage{titling}
\posttitle{\par\end{center}}
\title{\textbf{Intrinsic energy conversion mechanism via telescopic extension and retraction of concentric carbon nanotubes}}

\usepackage[affil-it]{authblk}

\date{}
\author[1]{Zhengrong Guo\thanks{zrguo@shu.edu.cn}}
\author[2]{Hongwei Zhang}
\author[2]{Jianxin Li}
\author[1]{Jiantao Leng}
\author[3]{Yingyan Zhang}
\author[1]{Tienchong Chang\thanks{tchang@staff.shu.edu.cn}}

\affil[1]{Shanghai Institute of Applied Mathematics and Mechanics, Shanghai Key Laboratory of Mechanics in Energy Engineering, Shanghai University, Shanghai 200072, People's Republic of China}

\affil[2]{State Key Laboratory of Ocean Engineering, School of Naval Architecture, Ocean and Civil Engineering, Shanghai Jiao Tong University, Shanghai 200240, People's Republic of China}

\affil[3]{School of Computing, Engineering and Mathematics, Western Sydney University, Penrith, NSW 2751, Australia\vspace{-4\baselineskip}}

\begin{document}
\maketitle

\begin{abstract}

The conversion of other forms of energy into mechanical work through the geometrical extension and retraction of nanomaterials has a wide variety of potential applications, including for mimicking biomotors. Here, using molecular dynamic simulations, we demonstrate that there exists an intrinsic energy conversion mechanism between thermal energy and mechanical work in the telescopic motions of double-walled carbon nanotubes (DWCNTs). A DWCNT can inherently convert heat into mechanical work in its telescopic extension process, while convert mechanical energy into heat in its telescopic retraction process. These two processes are thermodynamically reversible. The underlying mechanism for this reversibility is that the entropy changes with the telescopic overlapping length of concentric individual tubes. We find also that the entropy effect enlarges with the decreasing intertube space of DWCNTs. As a result, the spontaneously telescopic motion of a condensed DWCNT can be switched to extrusion by rising the system temperature above a critical value. These findings are important for fundamentally understanding the mechanical behavior of concentric nanotubes, and may have general implications in the application of DWCNTs as linear motors in nanodevices. 

\end{abstract}

\begin{multicols}{2}

Designing small motors to perform mechanical tasks at the microscale is one of the most exciting challenges in nanotechnology. There has been countless natural molecular motors, namely biomotors, with remarkable performance inspiring people to develop man-made nanomotors \cite{Wang2009}. A compact and elegant type of biomotor \cite{Theriot2000}, such as actin filament or protein microtubule, is a quasi-one dimensional structure where the mechanical work is produced by the geometrical extension and retraction. The force and motion generated during the extension and retraction plays an essential role in many biological processes, ranging from muscle contraction to cell division \cite{Inoue1995}. However, whether such structural extension and retraction can be used in a man-made nanomotor to do mechanical work is yet to debate, although man-made nanomotors mimicking other types of biomotor, e.g, rotary biomotor \cite{Eelkema2006}, have been extensively demonstrated  \cite{Hanggi2009,Fennimore2003,Bailey2008,Chang2010,Chang2015,Barzegar2016}.

Multi-walled carbon nanotubes (MWCNT), composed of concentric cylindrical graphene layers, is one of the most striking synthetic nanomaterials whose coaxial structures can be easily elongated and shrunk back via interlayer sliding \cite{Baughman1999,Yu2000,Moore2015}. The strong intralayer $sp^2$ chemical bonds make each individual tube of MWCNTs the strongest structures ever synthesized \cite{Treacy1996}. In contrast, the interlayer interaction of MWCNT is predominantly van der Waals (vdW) force, leading to ultra-smooth contacts between the adjacent layers \cite{Kolmogorov2000}. In a pioneering experimental study \cite{Cumings2000}, Cumings and Zettl demonstrated that the inner tube of a MWCNT can be pulled out by a small force applied on a carbon-tipped manipulator contacting with one end of the inner tube, and it can also be pulled back to its initial position by the vdW force after being released. Such extension-retraction cycle does not cause structural damage on the adjacent surfaces \cite{Kis2006} and thus could be operated repeatedly. This telescopic extension and retraction involves energy conversions among mechanical work, vdW potential and kinetic energy, all of which, however, are in essence mechanical energy. Whether energy in other forms, such as thermal energy, can be converted into mechanical work through the telescopic extension and retraction is virtually unknown. In fact, due to its extremely importance for nanoscale energy conversion, the posibility of using MWCNTs to convert thermal energy into mechanical work has been examined in a few studies\cite{Schoen2006,Barreiro2008,Guo2011,Guo2012}. For example, a thermophoretic motor proposed by Schoen et al. \cite{Schoen2006} (later demonstrated experimentally by Barreiro et al. \cite{Barreiro2008}) showed that nanoscale linear motions can be induced by a thermal gradient. However, no geometrical extension and retraction is accompanied with thermal-to-mechanical energy conversion in these studies.

Here, we reveal that there in fact exists an intrinsic energy conversion mechanism between thermal energy and mechanical work through telescopic motions of MWCNTs. A telescopic extension of a MWCNT inherently converts thermal energy into mechanical work, while a telescopic retraction converts mechanical energy into heat. Meanwhile, the energy dissipation in these two conversions are expected to be very small because the two processes are thermodynamically reversible. This implies that the MWCNTs may have a great potential to mimic biomotors.

\begin{figure*}[htb]
\centering \includegraphics[width=0.85\textwidth]{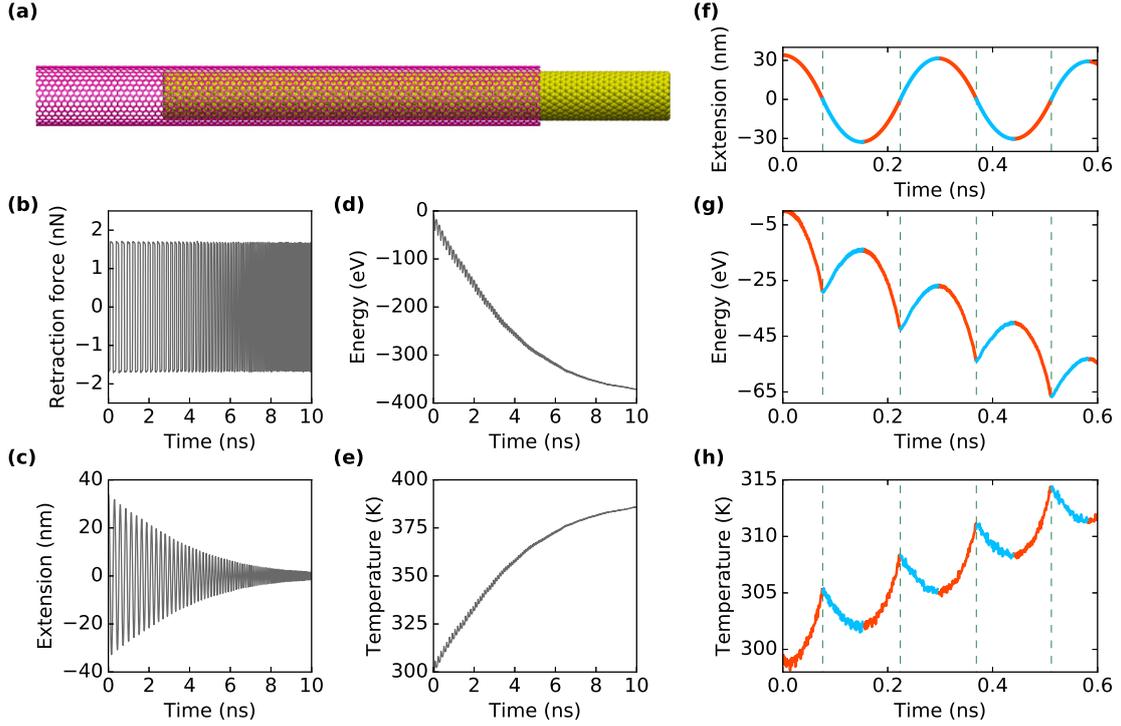}
\caption{(\textbf{a}) Schematic of a (17,17)/(21,0) DWCNT oscillator. (\textbf{b})-(\textbf{e}) Variation of retraction forces, distance of telescopic extension, oscillation energy and system temperature with respect to simulation time. (\textbf{f})-(\textbf{h}) A closer view of the extension distance, oscillation energy and system temperature from the starting time to 0.6 ns.}
\end{figure*}

This energy conversion mechanism is extensively explored by performing molecular dynamic simulation on double-walled carbon nanotube (DWCNT), since the DWCNT is the simplest system to investigate the interaction between concentric tubes in MWCNTs \cite{Moore2015}. Our simulation is carried out using a classical molecular dynamics package (LAMMPS) \cite{Plimpton1995} with a time step of 1 fs. The intramolecular interaction inside the tubes is described by a reactive empirical bond order (REBO) potential \cite{Brenner2002} and the interlayer interaction is described by a Lennard-Jones 12-6 potential ($\epsilon=2.968\ \mathrm{meV},\ \sigma=0.307\ \mathrm{nm}$).

In terms of the interlayer vdW potential energy, a DWCNT reaches to its equilibrium state when its inner and outer tubes overlap fully with each other (i.e. inner and outer tubes have the same length as shown in Figure 1a). At the equilibrium position, an initial perturbation will trigger the telescopic oscillation of the inner tube from both ends of the outer tube. This telescopic oscillation model, proposed by Zheng et al. \cite{Zheng2002}, has been widely used to investigate the friction between concentric tubes \cite{Guo2003, Rivera2003, Zhao2003, Legoas2003, Legoas2004, Rivera2005}. Here we use this model to monitor the energy conversion during the telescopic motions.

The simulated system is a (17,17)/(21,0) DWCNT with a  length of 35 nm. The oscillation starts at an initial extension distance of 34 nm. To better observe the oscillation, we restrain the translational motion of the outer tube (by fixing its two ends) and the rotation of the inner tube about the common axis. The simulation is performed under an adiabatic condition (using an NVE ensemble) at an initial temperature of 300 K. The general features of the oscillation obsevered here are consistent with the previous simulation studies on DWCNT oscillators \cite{Guo2003, Rivera2003, Zhao2003, Legoas2003, Legoas2004, Rivera2005,Cai2014}. The retraction force and the extension distance changing with time are shown in Figure 1(b) and 1(c), respectively. Owing to the friction between the relative tubes, the oscillation is damped over time. The oscillation energy (mechanical energy in oscillation, consisting of the kinetic energy of the inner tube and the interlayer vdW potential energy) decreases over time as shown in Figure 1(d), while the system temperature (in which the bias from mass center velocity of the inner tube has been subtracted) increases with time as shown in Figure 1(e).

However, by taking a closer look at Figures 1(d) and 1(e), we notice an intriguing phenomenon, i.e. the system temperature and the oscillation energy changes not solely with time in response to the frictional dissipation, but also varies periodically with the telescopic extending and retracting positions of the inner tube, as clearly shown in Figures 2(f-h). The temperature increases during each retraction process and decreases during each extension process. In a single oscillation cycle, temperature reaches a local maximum at the full overlapping position and goes to a local minimum at the largest extension position. In contrast, the oscillation energy increases during each extension process and decreases during each retraction process. The maximum and minimum oscillation energy values are obtained at the largest extension position and the full overlapping position, respectively. Since the simulation is performed under an adiabatic condition, we can safely conclude that the reduced oscillation energy in a retraction process is converted into thermal energy somewhat, thus leading to an increase in the system temperature. Likewise, the thermal energy reduction associated with the temperature decrease is converted to oscillation energy in an extension process.

\begin{figure}[H]
\includegraphics[width=0.485\textwidth]{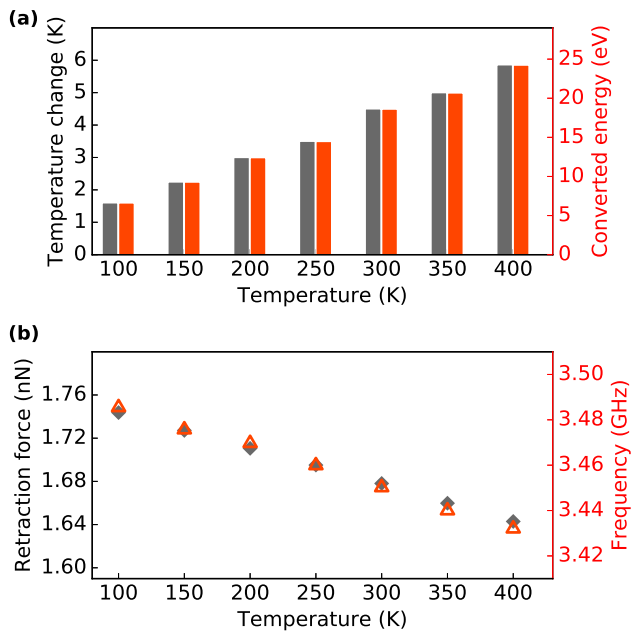}
\caption{Adiabatic simulations of a (17,17)/(21,0) DWCNT oscillator with different initial temperatures.  (\textbf{a}) Temperature change and converted energy versus system temperature. (\textbf{b}) Retraction force and frequency of oscillation versus system temperature.}
\end{figure}

Since the energy conversion is between thermal and mechanical forms, the system temperature may also play an important role in the oscillation. To confirm this point, we further simulate the same oscillator at different initial temperatures, ranging from 100 K to 400 K. Figure 2 shows the amount of converted energy and the temperature change in an extension as functions of the initial temperature, as well as the retraction force and the oscillation frequency at the beginning of the simulation. It is readily seen from Figure 2 that all these quantities change approximately linearly with the system temperature. The temperature change and the converted energy increase with the system temperature (Figure 2a). In contrast, the retraction force decreases with the increase of the system temperature (Figure 2b). Consequently, the decreasing retraction force leads to the decrease of the frequency of oscillation with respect to the system temperature. Interestingly, although no special attention has been given to these temperature-dependent phenomena, they indeed had appeared in some previous studies on the DWCNT oscillator \cite{Zhao2003,Guo2003}. For example, a figure in a research by Guo et al. \cite{Guo2003} on the study of friction shows that the oscillation has a higher initial frequency at a higher starting temperature.

To gain a clearer picture of the energy conversion processes, we simulate the system under isothermal conditions (NVT) using a Berendsen isothermal bath to keep the system temperature constant at 300 K. We extend and retract the inner tube from/into the outer tube at constant speeds (100-500 m/s), following the experimental approach to manipulate extensions and retractions of MWCNTs \cite{Cumings2000, Kis2006}. Figure 3 shows the cumulative heat exchanges between the system and the isothermal bath during the simulation. The heat transfers from the isothermal bath to DWCNT during the extension processes and inversely during the retraction processes. In other words, the extension processes are endothermic processes, and the retraction processes are exothermic processes. Note that there is a slight delay of the heat transfer with respect to telescopic position, which is caused by the high extending and retracting speeds as well as the relatively low heat transfer rate between the isothermal bath and the system.

By the end of an extension-retraction cycle, the cumulative heat exchange is positive as shown in Figure 3(c),  signifying heat dissipation. However, the amount of this dissipative heat decreases almost linearly with the decrease of the sliding speed. It means a negligible dissipative heat can be expected when the extension-retraction speed slows down to near zero. In other words, the energy conversions in an extension process and a retraction process are in essence thermodynamically reversible from one to another. The fundamental cause for this reversibility is that the entropy of system (so as the thermodynamic state) is determined by the telescopic extension distance. It is worth noting that the dissipative heat is most likely generated by the frictional dissipation because the friction in those systems is known to be linearly dependent on the sliding velocity \cite{Servantie2003, Tangney2004, Zhang2016}.

\begin{figure}[H]
\includegraphics[width=0.485\textwidth]{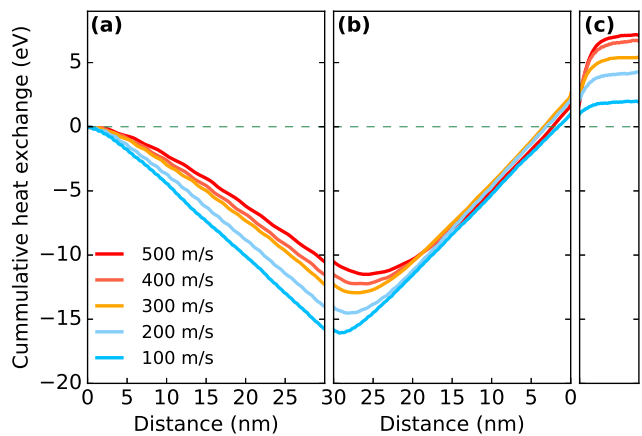}
\caption{Isothermal simulations of an extension-retraction cycle of a (17,17)/(21,0) DWCNT.  Cumulative heat exchange between the DWCNT and an isothermal bath during the extension process (\textbf{a}), the retraction process (\textbf{b}) and a successive system equilibration (\textbf{c}) for 100 ps.}
\end{figure}

How does a telescopic extension or retraction give rise to the observed energy conversion between heat and mechanical work? A well-known similar phenomenon is the energy conversion in a Stirling engine in which the mechanical work is achieved by the expansion of the contained gas in a cylinder. The underlying mechanism is the thermodynamic tendency of the gas to increase its configurational entropy (or the tendency of the gas to fill the available space). A previous research had shown that the detachment of a graphene layer from adhesion state to freestanding state would increase the configurational entropy of the graphene layer since the freestanding one has less constraint \cite{Guo2015}. The interaction between concentric tubes apparently has a similar effect on configurational entropy of the system. Take the telescopic extension process in the adiabatic oscillation as an example, the thermodynamic entropy of the system decreases as the system temperature decreases. To keep the total entropy constant or growth according to the second law of thermodynamics, the configurational entropy of concentric tubes must increase with the decrease of overlapping length in the extension. In contrast, the entropy of system decreases with the increase of overlapping length in a telescopic retraction. This mechanism can explicitly explain why the oscillation quantities depend linearly on temperature. The converted energy $\Delta E$ between heat and mechanical energy due to a configuration entropy change $\Delta S$ is theoretically expressed as $\Delta E=T \cdot \Delta S$, where $T$ is the temperature. This is why the temperature change as well as the energy conversion linearly increase with the system temperature (Figure 2a). Meanwhile, from a force perspective, the configuration entropy variation with respect to a moving path will give rise to a thermodynamic force, namely, entropic force. This entropic force, such as the pressure generated by gas on a surface, reflects the thermodynamical tendency of a system to increase its entropy. Compared with a static force, such as a vdW force, an entropic force is repulsive and inherently dependent on the temperature. The entropic force toward a moving path \emph{x} is generally expressed as $ F_{\mathrm{entr}} = -T \cdot \nabla_{x} S$. Here, $x$ is the distance of the telescopic extension, and the retraction force in fact consist of an attractive vdW force and a repulsive entropic force, 
\begin{equation}
F_{\mathrm{retr}}=F_{\mathrm{vdW}}-T \cdot \nabla_{x} S.
\end{equation}
Therefore, the retraction force decreases linearly with rising temperature due to the entropic force, as evidenced by the results in Figure 2(b). Interestingly, this entropic force has been observed by some studies in investigation of thermophoresis in DWCNT \cite{Guo2011,Guo2012,Li2015}, where direct calculation of the forces on carbon atoms showed that the entropic force is acting on the edge atoms of the outer tube and proportional to the system temperature \cite{Guo2011}. These researches demonstrated that the entropic force (or edge force) plays a dominant role in the thermophoresis in MWCNTs \cite{Guo2011,Guo2012}.

To further evaluate the important role the entropic force plays in the telescopic motion of DWCNTs, we calculate the entropic forces and vdW forces for two sets of DWCNTs, (17,17) and (29,0) outer tubes with various inner tubes (\emph{m},\emph{n}) \cite{Note1}. The simulations are performed under isothermal conditions. The vdW forces are calculated from the retraction forces with the system temperature of 0.1 K, at which the entropic forces are negligible. The entropic forces are calculated by the subtraction of the retraction forces (at 300 K) from vdW forces. We find that both vdW force and entropic force depend heavily on the intertube spaces as shown in Figure 4. It is clearly seen that an initial decrease of the intertube space from 0.35 nm to 0.34 nm lead to a slight increase of the vdW force. However, with the further decrease of intertube space, the vdW force decreases. The maximum vdW force is found at an intertube space of around 0.34 nm in accord with the minimum vdW energy, indicating that 0.34 nm is the equilibrium intertube space of DWCNTs. Condensing the intertube space further from 0.34 nm compresses the inner and outer tubes in the radial direction and enhances the system potential energy \cite{Saito2001,Zhang2007}. As a consequence, the vdW force drops. In contrast, the entropic force increases with the decrease of intertube space because a closer proximity of the adjacent tubes imposes stronger constraint upon the thermal vibration of atoms \cite{He2006}. Note that the intertube spaces of DWCNTs under consideration are within the allowable range (i.e. 0.27-0.42nm) determined by experimental observations \cite{Kharissova2014}.

\begin{figure}[H]
\centering
\includegraphics[width=0.485\textwidth]{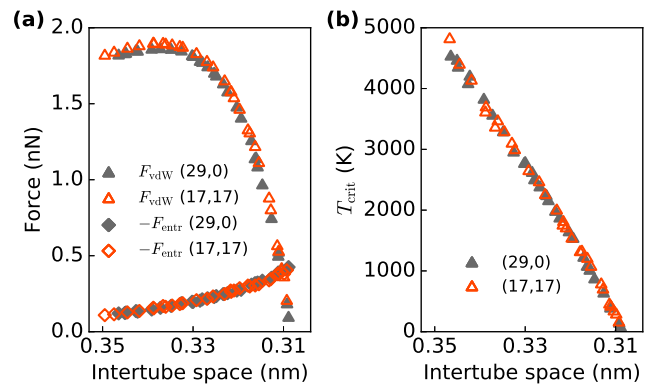}
\caption{(\textbf{a}) Entropic force and vdW force versus intertube space at 300 K. Each value corresponds to a (17,17)/(\emph{m},\emph{n}) or (29,0)/(\emph{m},\emph{n}) DWCNT. (\textbf{b}) Critical temperature of (17,17)/(\emph{m},\emph{n}) and (29,0)/(\emph{m},\emph{n}) DWCNTs as functions of intertube space.}
\end{figure}

\begin{figure*}[]
\includegraphics[width=1\textwidth]{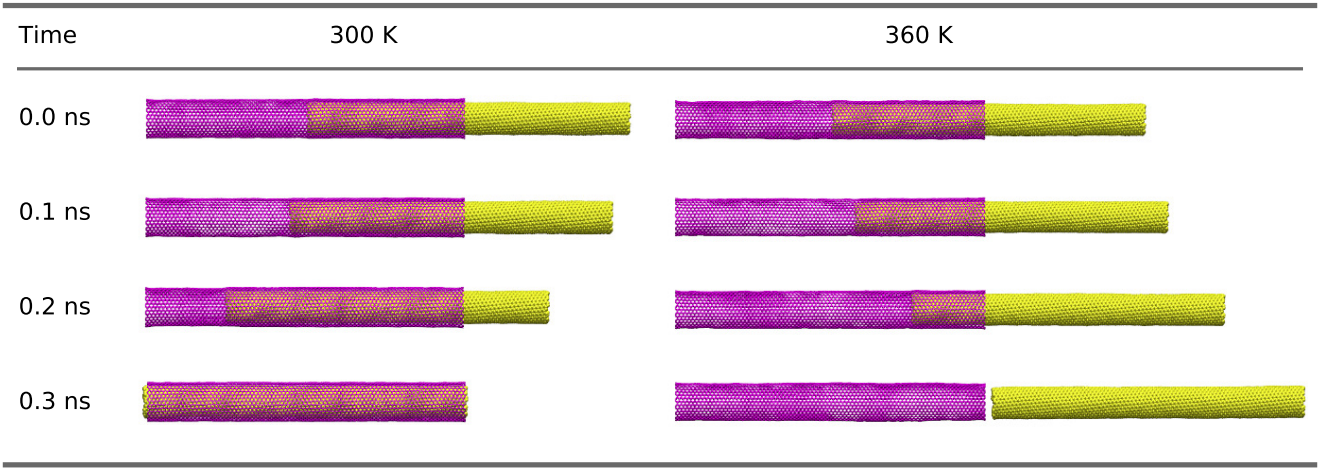}
\caption{Spontaneously telescopic motions of a half-extruded (17, 17)/(15, 11) DWCNT at 300 K and 360 K.}
\end{figure*}

At an intertube space of 0.34 nm, the vdW force is about one order of magnitude larger than the entropic force. Therefore the spontaneously telescopic motions in such DWCNT favors retraction as had usually been observed in experiments \cite{Cumings2000}. However, for a DWCNT with an intertube space of around 0.31 nm, the entropic force is close to the vdW force. This is remarkable because the spontaneously telescopic motions will be determined by the competition of these two forces instead of by the vdW force solely. The entropic force is temperature-dependent, while the vdW force is constant, independent of temperature. There must therefore exist a critical temperature $T_{\mathrm{crit}}$, at which the two forces are equal in magnitude and thus the net retraction force is zero. When the system temperature is below $T_{\mathrm {crit}}$, spontaneously telescopic motion would still be retraction favored. However, when the temperature is higher than $T_{\mathrm {crit}}$, the total retraction force would become repulsive and the spontaneous telescopic motion turns to be extension. To verify such phenomenon, we simulate a half-extruded (17,17)/(15,11) DWCNT whose intertube space is 0.31 nm and the expected $T_{\mathrm{crit}}=330$ K (by $T_{\mathrm{crit}}=F_{\mathrm{vdW}}/\nabla_{x} S$). Figure 5 illustrates the typical snapshots of spontaneous motions of the DWCNT at 300 K and 360 K, respectively. At room temperature (300 K), the telescopic retraction happens, while at 360 K, the telescopic extension takes place.

The critical temperature is highly sensitive to and linearly dependent on the intertube space, as shown in Figure 4(b). A small increase in the intertube space from 0.31 nm to 0.32 nm can lead to a remarkable jump in the critical temperature from 300 K to 1800 K, beyond which MWCNTs are no longer thermodynamically stable \cite{Liew2005,Sarkar2013}. For a DWCNT with its intertube space being the equilibrium distance of 0.34 nm, the critical temperature is estimated to be 4000 K. It means that in allowable temperature range those DWCNTs always favor retraction. In contrast, in the cases that intertube space is smaller than 0.305 nm, the vdW forces become repulsive, and the telescopic extension is always favored even when temperature is at 0 K. However, since the possible combination of inner-outer tubes is infinite theoretically, thereby it is always possible to find a DWCNT with any desired critical temperatures.

From a technological perspective, the entropic forces in a DWCNT fall within a useful and accessible range. For instance, the entropic force in a DWCNT at room temperature is about 150 pN, which is close to the driving force generated by a typical protein biomotor \cite{Wang2009}. As a one-dimensional structure, DWCNT is quite feasible to be integrated with other nanostructures and the telescopic motions can be achieved easily. Thus it is promising for the MWCNTs to be utilized as actuators in nanomachines. Another possible application of this driving mechanism is to array a large number of DWCNTs to form a macroscale fiber with its driving force depending on the containing number of MWCNTs \cite{Baughman2002}.  In particular, a millimeter-scale fiber consisting of (17,17)/(15,11) DWCNTs could generate a driving force on the order of tens of Newton while a centimeter-scale one could generate a driving force on the order of thousands of Newton. Thereby, such a muscle-like fiber can be used in a wide variety of situations.

In conclusion, we have shown an intrinsic energy conversion mechanism between heat and mechanical work via the telescopic extension and retraction of MWCNTs. The converted energy from telescopic motion and the retraction force between nested nanotubes depends on the system temperature. The underlying mechanism for this temperature-dependent energy conversion is the entropy change related to the overlapping length of concentric tubes. We further found a temperature-tunable spontaneously telescopic motion in the condensed DWCNTs. Furthermore, the energy conversion in the telescopic motions is thermodynamically reversible; the future devices based on this mechanism thus can perform their function in an energy efficient way. Our findings suggest that the intrinsic energy conversion mechanism in concentric MWCNTs hold great promise in the practical applications in future mechanical nanodevices, such as thermal motors, temperature-controlled switches and artificial muscles.

\section*{Acknowledgement}

The authors from Shanghai University acknowledge the financial support from NSFC (Grant Nos. 11602132, 11425209). The molecular dynamic simulations were carried out on the computing platform of the International Center for Applied Mechanics in Energy Engineering (ICAMEE), Shanghai University. 


\label{Bibliography}
\bibliographystyle{unsrtnat}


\end{multicols}
\end{document}